\documentclass[preprint]{revtex4}
\usepackage{psfig}
\usepackage{graphicx}%
\usepackage{dcolumn}
\usepackage{amsmath}

\makeatletter
\def\btt#1{\texttt{\@backslashchar#1}}%
\DeclareRobustCommand\bblash{\btt{\@backslashchar}}%
\makeatother

\begin{document}
\title{Relating Quantum Information to Charged
Black Holes}

\author{Xian-Hui Ge$^{1,2,3}$}\email{gexh@shao.ac.cn}

\author{You-Gen Shen$^{1,2,4}$}\email{ygshen@shao.ac.cn}
\affiliation{$^{1}$ Shanghai Astronomical Observatory, Chinese
Academy of
Sciences, Shanghai 200030, China.\\
$^{2}$ National Astronomical Observatories, Chinese Academy of
Sciences,
Beijing 100012, China.\\
$^{3}$Graduate School of Chinese Academy of Sciences, Beijing
100039, China\\
$^{4}$ Institute of Theoretical physics, Chinese Academy of
Sciences, Beijing 100080, China.
}%

\begin{abstract}
Quantum non-cloning theorem and a thought experiment are discussed
for charged black holes whose global structure exhibits an event
and a Cauchy horizon. We take  Reissner-Norstr\"{o}m black holes
and two-dimensional dilaton black holes as concrete examples. The
results show that the quantum non-cloning theorem and the black
hole complementarity are far from consistent inside the inner
horizon. The relevance of this work to non-local measurements is
briefly discussed.
\\
\textbf{Keywords}: quantum non-cloning theorem, black hole
information, black hole
complementarity\\
{PACS numbers: 03.67.-a, 03.65.Ud, 04.62.+v,
04.70.Dy }

\end{abstract}

\date{\today}


\maketitle

\section{Introduction}The loss of information of a black hole has long
been  an interesting question
 in the theoretical physics[1]. Although many physicists have devoted time searching for the answer to
 this question[2-8], the fundamental
 solution is still elusive. On the other hand,
 the development of quantum information theory
[9 ] draws attention to fundamental questions about what is
physically possible and what is not. For instance, the quantum
non-cloning theorem [10], which asserts, unknown pure states can
not be reproduced or copied by any physical means. Recently, there
are growing interests in the quantum non-cloning theorem. The
original proof of this theorem [10] shows that the cloning machine
violates the quantum superposition principle. The second version
of the quantum non-cloning theorem states that a violation of
unitarity makes cloning two non-orthogonal states
impossible[11,12]. However, the third version argues that if the
unitarity of two non-orthogonal states is destroyed and only if
they are linearly independent, then the states which are secretly
chosen from a certain set can be probabilistically cloned [13,14].
 In this paper, we wish to discuss what the quantum non-cloning theorem might
infer to a charged
black hole.\\
  \hspace*{7.5mm}We notice that, among the many efforts endeavoring to resolve the
  black hole information paradox, Susskind, Thorlacius and Uglum suggest
  the black hole complementarity principle [3] which can be formulated as follows:
  i) From the point of view of an external observer the region
  just outside the horizon, stretched horizon, acts like a very
  hot membrane which absorbs, thermalizes and emits any information
  that falls to the black hole;
  ii) From the point of view of a freely falling observer there is
  nothing special at the horizon so a freely falling observer can
  cross the horizon in his way to the singularity.\\
    \hspace*{7.5mm} As it is pointed out by Lowe et.al[5],
    this principle describes physical
    pictures which are apparently contradictory. According to
    this principle, an observer who remains
    outside the event horizon of a black hole can describe the
    black hole as a hot membrane, the stretched horizon, which
    absorbs and stores anything falling onto it. And there is no information loss: all the
    information stored in the membrane will eventually re-emitted
    in the form of Hawking radiation. On the other hand, an
    observer who falls freely into the black hole sees things very
    differently, no membrane, no stretched horizon, and nothing
    irregular at the event horizon. Moreover, from the point of
    view of the freely falling observer all the information entering into
    the black hole will never come back.
    Therefore, it seems that the black hole can
     act as a cloning machine because the matter which has fallen past the event horizon
     and the Hawking radiation are not different objects. If the re-emitted quantum information has a
     chance to fall into the black hole, then once it crosses the event horizon
     there will be definitely two copies of the same quantum information.
   This violates the basic principle of the quantum mechanics. In relation to this conflict,
     Susskind et al
     show that duplicate information will be never detected
     in a Schwarzschild black hole because any measurement inside the  black hole
     will require the energy far beyond the total energy of the black hole[4,15]. And if we make an assumption that
      quantum mechanics forbids information  cloning as meaning that no real observer is ever
 allowed to detect duplicate information, then the quantum non-cloning theorem is preserved. \\
 \hspace*{7.5mm}Nevertheless, we find that the relationship
 between the quantum non-cloning theorem and the black hole
 complementarity is still far from consistent for charged black
 holes whose global structure exhibits an event and a Cauchy horizon. In this paper, we wish to
discuss the quantum non-cloning theorem for Reissner-Norstr\"{o}m
black holes(RN) and two-dimensional (2D) dilaton black holes as
concrete examples, because the properties of RN black holes and 2D
dilaton black holes are different very much from Schwarzschild
black holes. This will be helpful for our thorough
understanding of the quantum non-cloning theorem and the black hole complementarity.\\
\section{A thought experiment conducted on an RN black hole}
  \hspace*{7.5mm}We first begin with the metric of general spherically symmetric
space-time, which can be written as[16,17] ( Planck units is
used:$\hbar=G=c=k=1$ hereafter)
\begin{equation}ds^{2} = e^{2U\left( {r} \right)}dt^{2} - e^{ - 2U\left( {r}
\right)}dr^{2} - R^{2}\left( {r} \right)\left( {d\theta ^{2} +
sin^{2}\theta d\varphi ^{2}} \right).
\end{equation}
Following the idea of [16], we can concentrate on the "near
horizon limit", and define $y=r-r_{H} $, where $r_{H}$ is the
event horizon and $y\ll r_{H}$. The metric
becomes\begin{equation}ds^{2} = e^{2U\left( {y} \right)}dt^{2} -
e^{ - 2U\left( {y} \right)}dy^{2} - R^{2}\left( {y}
\right)d\Omega^2.
\end{equation}
If we set\begin{eqnarray}\rho=\int e^{ - U\left( {y} \right)}dy,
 \omega=e^{U(y)} t/ {\int e^{-U(y)}dy},\end{eqnarray}
 then the metric has the form \begin{equation}
 ds^2=\rho^2 d\omega^2-d\rho^2-R^{2}\left( {y} \right)d\Omega^2.\end{equation}
 We further define \begin{equation}X^{+}=\rho
 e^{\omega},
 X^{-}=-\rho e^{-\omega},\end{equation} then the metric becomes \begin{equation}
 ds^2=dX^{+}dX^{-}-R^{2}\left( {y} \right)d\Omega^2.\end{equation}In this way, the
 horizon is no longer singular.
 In the following, we would like to investigate the quantum non-cloning theorem
 for RN black
 holes.\begin{figure}
\psfig{file=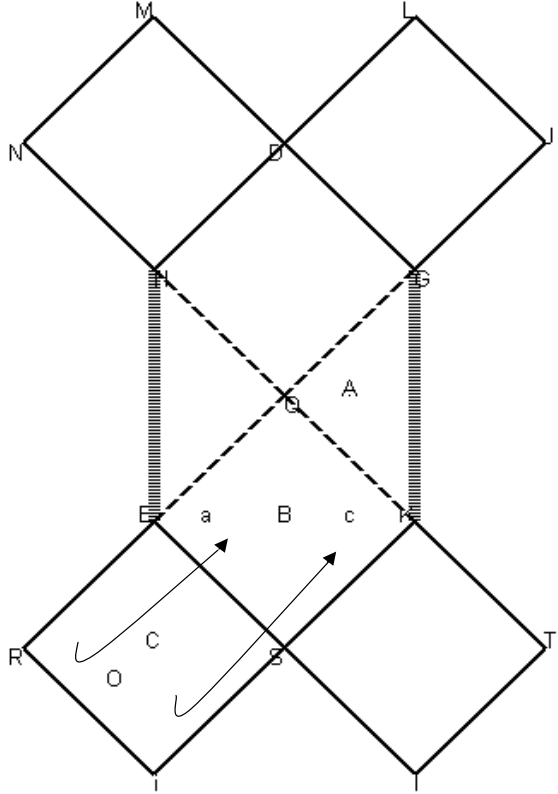,height=7in,width=5in}\caption{ A thought
experiment conducted on a RN black hole}
\end{figure}
The line elements of a RN black hole are given by
 \begin{equation}ds^2=(1-\frac{2M}{r}+\frac{Q^2}{r^2})dt^2-(1-\frac{2M}{r}
 +\frac{Q^2}{r^2})^{-1}dr^2-r^2(d\theta+sin\theta d\varphi)^2,\end{equation} where $M, Q$
  are the mass and the charge of the black
 hole. The horizon equation is \begin{equation}r^2-2Mr+Q^2=0,\end{equation} and the
 solutions
 \begin{equation}r_{H}=M+\sqrt{M^2-Q^2},  r_{I}=M-\sqrt{M^2-Q^2}\end{equation}
 are the event horizon and the inner horizon
 respectively. Thus the space-time can be distinguished into three regions:\\
 \hspace*{20mm}A: $0<r<r_{I}$;
  B: $r_{I}<r<r_{H}$;
  and  C: $r>r_{H}$.\\ We can find
  that, while the surface $r=r_{H}$ is an event
 horizon in the same sense that $r=2M$ is an event horizon in the Schwarzschild space-time,
 the surface $r=r_{I}$ is a horizon in a different sense. The original
 Schwarzschild singularity $r=0$ here has become the inner horizon, while the RN singularity $r=0$
 corresponds to a negative value of $r$ in the original Schwarzschild metric.
 In the following analysis, we still concentrate on the "near horizon limit" and consider a small angular
  region near a point on the
 horizon. Define
 \begin{eqnarray}&&y=r-r_{H};  y\ll r_{H},\nonumber\\
 &&\rho=\frac{M+\sqrt{M^2-Q^2}}{\sqrt[4]{M^2-Q^2}} (2y)^{\frac{1}{2}},\nonumber\\
 &&\omega=\frac{\sqrt{M^2-Q^2}}{(M+\sqrt{M^2-Q^2})^2}t,\end{eqnarray} and
 \begin{equation}X^{+}=\rho e^{\omega},  X^{-}=-\rho e^{-\omega}.\end{equation} Thus the
 metric can finally be written as
 \begin{equation}ds^2=dX^{+}dX^{-}-dx^{i}dx^{i} \end{equation}
 The lifetime of information stored in the stretched horizon is
called the black hole information retention time[4]. According to
the black hole complementarity, when a q-bit information is thrown
into a black hole, an observer outside the event horizon can
calculate how long it will be re-emmited in form of Hawking
radiation. We can derive the information retention time of the RN
black holes as follows.
  When a RN black hole radiates,
 the spectrum of particles is given by the Planck distribution[18]:
 \begin{equation}d E_{\omega}=\frac{(\omega-e\phi)^3 d\omega}{e^{(\omega-e\phi)/T_{H}}-1},
 \end{equation}where $dE_{\omega}$ is the radiation energy in the spectral
 range $\omega$ to $\omega+d\omega$.
 Integrating over $\omega$ from $e\phi$ to $\infty$, one obtains the rate at
 which the black hole loses
 energy, ie, mass
 \begin{equation}\frac{dM}{dt}=-\sigma T_{H}^{4}A.\end{equation}
 $A=4\pi{ r_{H}}^2$ is the area of the event horizon and $\sigma$
 the Stefan-Boltzman constant.
 Thus, for RN black holes with $T_{H}=\frac{(r_{H}-r_{I})}{4\pi r_{H}^2}$, it is given by
 \begin{equation}\frac{dM}{dt}=-\frac{\sigma (r_{H}-r_{I})^4}{(4\pi)^3 r_{H}^6}.
 \end{equation} We integrate (15) to get
 \begin{equation}\int_{0}^{t} dt=-\frac{(4\pi)^3}{\sigma}\int_{M,Q}^{0,0}
 \frac{r_{H}^6 dM}{(r_{H}-r_{I})^4}.\end{equation}
Substituting (9) into (16)and assuming $Q=\lambda M$, where
$0\leq\lambda<1$, we have
\begin{equation}t=\frac{(4\pi)^3}{3\sigma}\frac{(1+\sqrt{1-\lambda})^6}
{16(1-\lambda)^2}M^3.\end{equation} Hence, the information
retention time should have the same order of magnitude, this is to
say $t_{R}\sim M^3$. Now, let's consider a thought experiment
which is originally considered in [4] and repeated later in the
review article[15]. For simplicity, we just repeat the same
experiment process of that in [4,15] for a RN black hole, for that
is helpful for our understanding on the relationship between black
hole complementarity and quantum non-cloning theorem. When a RN
black hole is formed,
 a q-bit
 information is thrown in before the black hole has a chance to
 evaporate. Here the information must be quantum information and has the general
 form: $\mid \varphi>=a \mid 0> + b\mid 1> $, because only quantum information
 has the nature of non-cloning. According to the observer who falls with the q-bit,
 the information at a later time will be localized behind the
 horizon at a point (\emph{a}); see figure1. On the other hand, an observer outside the
 horizon eventually sees all of the energy returned in the form of
 Hawking radiation. Thus, according to the black hole complementarity,
 a measurement can be performed on the radiation
 and the original information can be determined.
  We assume there is an observer \emph{O} stationed  outside the the
horizon to collects information stored in the infalling q-bit. At
that time, the observer jumps into the black hole, carrying the
information to point(\emph{c}) behind the horizon. Now there are
two copies of the q-bit behind the horizon one at(\emph{a}) and
one at (\emph{c}). A signal from (\emph{a}) to (\emph{c}) can
reveal that information has been duplicated and then the quantum
non-cloning theorem is violated. The analysis as follows can show
us how the quantum non-cloning theorem is preserved
in region B and how it is violated in region A. \\
 \hspace*{7.5mm}In the thought experiment, the point (\emph{c}) must occur before
the trajectory of \emph{O} intersects the inner horizon. On the
other hand, \emph{O} may not cross the event horizon until the
information retention time has elapsed. The implication of the two
constraints is most easily seen using the following
coordinates:\begin{eqnarray}&&X^{+}=\rho
e^{\omega},\nonumber\\&&X^{-}=-\rho e^{-\omega},\nonumber\\
&&\omega=\frac{\sqrt{M^2-Q^2}}{(M+\sqrt{M^2-Q^2})^2}t=\frac{\sqrt{1-\lambda}
 t}{(1+\sqrt{1-\lambda})^2 M}.\end{eqnarray}
 An observer outside the horizon must wait a time $t \sim CM^3$
 (the time which we have obtained in (17) and here $C$ is a positive constant), to
 collect a bit from the Hawking radiation. Thus, the observer may not jump into the
 black hole until $X^{+}\sim e^{CM^2}$, which
means that \emph{O} should be at a point satisfying
 \begin{equation}X^{-}< e^{-CM^2}.\end{equation}This requires that the message
 sent between (\emph{a}) and(\emph{c})
 must be sent within a time interval $\delta t$ of the same order of magnitude,
 an incredibly short time ($\delta t\sim e^{-CM^2}$). \\
\hspace*{7.5mm} The uncertainty principle of quantum mechanics
requires that\begin{figure}
\psfig{file=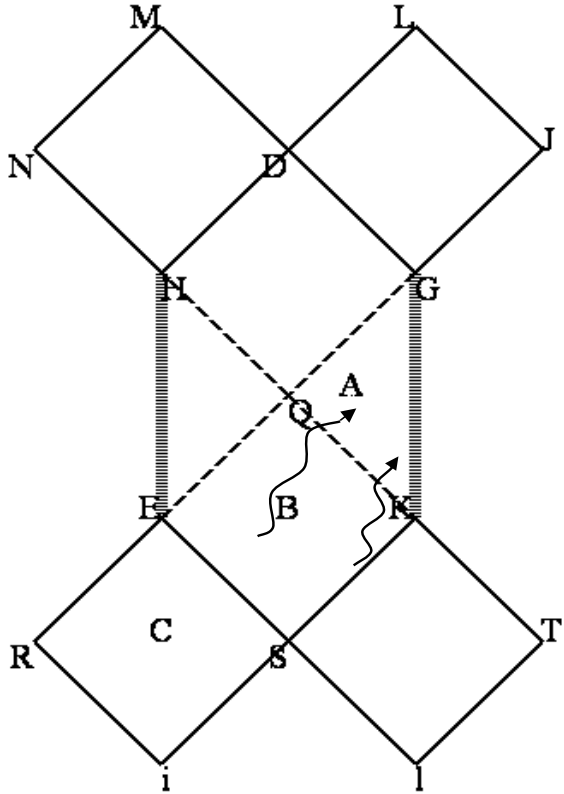,height=7in,width=5in}\caption{A thought
experiment conducted near the inner horizon}
\end{figure}
 the quanta of the message have energy of
 order $(\delta t)^{-1}$, which can be written as
 \begin{equation}E_{signal}\sim e^{CM^2}.\end{equation}Clearly, the energy
 required is much greater than the total energy
  or mass of the black hole. Therefore, it is impossible to detect duplicate
  information in the region between the event horizon and the
  inner horizon of RN black holes. The physical process discussed
  in [4,15] also show that any communication between (\emph{a}) and(\emph{c}) need a
  super-planckian frequency. As it is pointed in [15],
   there must be something wrong with the
  usual ideas of local quantum field theory in black hole background because
  a theory should not predict things which are in
  principle unobservable.\\
\hspace*{7.5mm}  In fact, when we turn to discuss
  the quantum non-cloning theorem within the inner horizon,
the duplicate information is not undetectable any longer. RN black
holes are different from Schwarzschild black holes: While
Schwarzschild black holes have space-like singularities, the
singularity of a RN black hole is time-like. As the inner horizon
is not a mathematical singularity of the geometry, we could
concentrate on the near horizon limit. We consider a small angular
region near a point on the horizon and
define:\begin{eqnarray} &&y'=r-r_{I},y'\ll r_{I}\nonumber\\
&&\rho'=\sqrt{2 y'}\frac{M-\sqrt{M^2-Q^2}}{\sqrt[4]{M^2-Q^2}}\nonumber\\
&&\omega'=\frac{\sqrt{M^2-Q^2}}{(M-\sqrt{M^2-Q^2})^2}t,
 \end{eqnarray}
and\begin{eqnarray}&&
X'^{+}=\rho'e^{-\omega'},\nonumber\\
&&X'^{-}=\rho'e^{\omega'},
\end{eqnarray}
where $X'^{+}$ should correspond to $X^{-}$ for consistency.
 Then the metric near the inner horizon can be finally written as
 \begin{equation}ds^2=dX^{+}dX^{-}.\end{equation}
For an observer inside the inner horizon, the two copies of q-bit
must appear at a time interval no less than the so called black
hole information retention time (see figure2), which is to say
\begin{equation}
X'^{-}>\rho'e^{\omega'}.\end{equation}

  Therefore, according to the quantum uncertainty principle,
  the energy required to send a message to
detect duplicate information inside the inner horizon, reads
\begin{equation}E_{signal}\sim e^{-CM^2},\end{equation} which is apparently permitted by the quantum
mechanics. Indeed, once the q-bit information and the observer
have a chance to cross the inner horizon, the quantum non-cloning
theorem seems unavoidable violated. We can easily see from (7)
  that
  \begin{eqnarray}
  &&g_{00}<0, g_{11}>0, g_{22}>0,g_{33}>0, (r>r_{H})\nonumber\\
  &&g_{00}>0,g_{11}<0,g_{22}>0,g_{33}>0, (r_{H}>r>r_{I})\nonumber\\
  &&g_{00}<0, g_{11}>0, g_{22}>0,g_{33}>0, (r<r_{I}).\end{eqnarray}
    The region between the event horizon and the
     inner horizon is a region which  r is the coordinate of time,
     and t is the coordinate of space. However, for regions $r<r_{I}$
     and $r>r_{H}$, space and time are not
     interchanged. An observer who crossed the inner horizon can
     act freely and the singularity is avoidable for the observer. This is a main
     deference between a RN black hole and a Schwarzschild black hole.
     Thus, the black hole complementarity and the the quantum
     non-cloning theorem become incompatible in region A.\\
      \hspace*{7.5mm}Moreover, what discussed in [5,15] about the quantum Xerox principle is almost
      entirely cannot be applied to extremal RN black holes. For extremal RN black holes, the two
      horizons coincide $r_{H}=r_{I}$. The event horizon becomes degenerate, with nonzero area but
       vanishing temperature. Theoretically, there exists two possibilities for the black hole complementarity
       and the quantum non-cloning theorem: one is that black hole complementarity or
       quantum non-cloning theorem is wrong; the other
       is that they are both correct but there is some physics in RN black holes still unknown.
       As to this puzzle, we want to  say that, instability of the inner horizon is not discussed
       here. Small external perturbations to the inner horizon are not strong enough to change
       the metric but if we consider quantum electrodynamics process in region B, the
       metric can be modified and it can be proved that the true space-time singularity is indeed
       created and no inner horizon formed then [17]. Now, we still lack evidence to prove that
       the experiment conducted on a RN black holes can generate enough perturbations to
       change the nature of the inner horizon. And also, when we talk about q-bit information here,
       we talk about it by only making local measurements. Quantum information theory is not
       a local theory. The
       essence of quantum information theory is entanglement and non-locality.
       In quantum theory one talks about communication
      between distant entangled states, where each state separately sees only
       random events and it is through entanglement which quantum information is
       transmitted. Local measurements cannot be an effective way
       to  tell what is true and what is not true in the above thought
      experiments. However, we do not mean to talk about
       quantum information theory and the instability of the inner horizon in detail.
       To answer the above question we should first
       know the nature of entanglement in the background of curved space-time. This is
       a rather difficult problem because quantum gravity is needed.
       \section{quantum non-cloning theorem in Two-dimensional Dilaton black holes }
       It is interesting to extend our discussion to
      2D dilaton black holes. The properties of
       2D dilaton black holes have been widely
       investigated in the past several years and it was found
       that many basic results of standard 4D black hole physics
       find their counterpart in the 2D case[19]. The metric of the
       2D dilaton black hole adopted is  as follows:
       \begin{equation}
       dS^2=N^2dt^2-(N^2)^{-1}dr^2,\end{equation}

       where\begin{equation}
       N^2=1-\frac{2M}{\Lambda}e^{-\Lambda r}+\frac{Q^2}{\Lambda^2}e^{-2\Lambda r}.\end{equation}
$M$ and $Q$ are, respectively, the mass and charge of a black
hole, and $\Lambda$ is the cosmological constant. The horizons of
the black hole are
\begin{equation}r_{\pm}=\Lambda^{-1}ln[\Lambda^{-1}(M\pm\sqrt{M^2-Q^2})].\end{equation}
And it is assumed that $Q<M$. The space-time can also be seperated
into three regions:A: $0<r<r_{-}$;
  B: $r_{-}<r<r_{+}$;
  and  C: $r>r_{+}$ . The Hawking temperature of the black hole is
\begin{equation}T_{+}=\frac{\Lambda (M^2-Q^2)^{1/2}}{2\pi(M+(M^2-Q^2)^{1/2})}\end{equation}
We can further define\begin{eqnarray}&& y=r-r_{+}; y\leq
r_{+}\nonumber\\
&&\rho=\frac{\sqrt{2}(e^{\Lambda
y}-1)^{1/2}}{\sqrt{(M^2-Q^2)^{1/2}[M+(M^2-Q^2)^{1/2}]}}\nonumber\\
&&\omega=\frac{(M^2-Q^2)^{1/2}\left[M+(M^2-Q^2)^{1/2}\right]}{\Lambda}t,\end{eqnarray}
and
\begin{eqnarray}&&
X^{+}=\rho e^{-\omega}, X^{-}=\rho e^{\omega}\end{eqnarray} Then,
the metric can be rewritten as
\begin{equation}dS^2=dX^{+}dX^{-}\end{equation}
The black hole information retention time can be obtained by using
the one-dimensional Stefan-Boltzman formula
\begin{equation}\frac{dM}{dt}=\frac{\pi^2{T_{+}}^2k_{B}^2}{6}A,\end{equation}
where $k_{B}$ is the Boltzman constant and A is the "area" of the
2D dilaton black hole. After integration of Eq.(34), we can
express the information retention time in a simple way, say
\begin{equation}t_{R}\sim C M,\end{equation} where C is a
constant and $Q=\lambda M$ ($0\leq\lambda<1$) is assumed in the
integration. Following the same process in section II, we find
that the message sent between the two observers in region B
requires the quanta of energy $E_{signal}\sim e^{CM^3}$, which is
different from Eq.(24), but still greater than the total energy of
the black hole. While inside the inner horizon, the duplicate
information is no longer undetectable. After the similar
calculation to section II, we obtain the energy required to send a
message to detect duplicate information inside the inner horizon,
which goes as $E_{signal}\sim e^{-CM^3}$. The results also show
that the non-cloning theorem can be violated for the 2D dilaton
black holes.
     \section{Conclusion}In summary, we have discussed the quantum
non-cloning theorem
    in RN black holes and 2D dilaton black holes. We find that the quantum non-cloning theorem can
    well established in the region between the inner horizon and
    event horizon, but to be violated inside the inner horizon.
   This seems to indicate that the black hole complementarity principle or
    the quantum non-cloning theorem is confront with a challenge and we may need some new physics to
    resolve the dilemma. Quantum information theory and the instability of the inner horizon are
    proposed as possible directions. We would like to investigate them in our future work.
    In addition, what we discussed about the RN black holes is also applicable to
    other charged black holes such as Garfinkle-Horne dilaton black holes  and Gibbons-Maeda dilaton black holes[20].
    \\
    \hspace*{7.5mm}Recently, Horowitz and Maldacena (HM) have
  proposed an alternative model of black hole evaporation by imposing a
  final state boundary condition at black hole singularities, to
  resolve the apparent contradiction between string theory and
  semiclassical arguments over whether black hole evaporation is
  unitary [21]. This model requires a specific final state at black
  hole singularity which is perfectly entangled between the
  collapsing matter and the incoming Hawking radiation. Thus, information in the black hole
  can be "teleported" out in the outgoing Hawking radiation and the quantum non-cloning theorem can be well
  preserved in the whole process. The
  proposal might shed light on the problems suffered by the black
  hole complementarity. However, Gottsman, Preskill, and later
Yurtsever and Hockney argued that the proposed constraint must
lead to nonlinear evolution of the initial quantum state, and one
cannot ensure the black hole final state to be maximally
entangled[22,23,24]. Giddings and Lippert pointed out that the HM
scenario depends on assuming inside and outside Hilbert spaces and
may conflict with the expectations for observations made by the
inside obervers[25]. In [26], we extend HM's proposal to Dirac
fields and find that if annihilation of the infalling positrons
and the collapsed electrons inside the horizon is
  considered, then the nonlinear evolution of collapsing quantum state can be
  avoided.
\begin{center}\textbf{ACKNOWLEDGEMENTS}\end{center}
 The work has been supported by the National Natural
Science Foundation of China under Grant No. 10273017.

\end{document}